\documentclass[runningheads]{llncs}
\usepackage{amsmath}
\usepackage[T1]{fontenc}
\usepackage{graphicx}
\usepackage{amssymb}
\usepackage{comment}
\usepackage{mathtools}
\usepackage{multirow}
\usepackage{booktabs}
\usepackage{hyperref}
\footnotesize
\tiny
\begin{document}
\title{Learned Image Compression for HE-stained Histopathological Images via Stain Deconvolution}
\titlerunning{Stain quantized WSI compression}
%
\author{Maximilian Fischer\inst{1,2} \and
Peter Neher\inst{1,2,11} \and
Tassilo Wald \inst{1} \and
Silvia Dias Almeida\inst{1,4} \and
Shuhan Xiao\inst{1,3} \and
Peter Sch\"uffler\inst{6} \and
Rickmer Braren\inst{7} \and
Michael G\"otz\inst{1,5} \and
Alexander Muckenhuber\inst{6} \and
Jens Kleesiek\inst{8,9} \and
Marco Nolden\inst{1,10} \and
Klaus Maier-Hein\inst{1,3,4,10,11} 
}
\authorrunning{Fischer et al.}

%
\institute{ Institute
Division of Medical Image Computing, German Cancer Research Center (DKFZ), Heidelberg, Germany\and
German Cancer Consortium (DKTK), partner site Heidelberg\and 
Faculty of Mathematics and Computer Science, Heidelberg University, Heidelberg, Germany\and
Medical Faculty, Heidelberg University, Heidelberg, Germany\and
Clinic of Diagnostics and Interventional Radiology, Section Experimental Radiology, Ulm University Medical Centre, Ulm, Germany\and
School of Medicine, Institute of Pathology, Technical University of Munich, Munich, Germany\and
Department of Diagnostic and Interventional Radiology, Faculty of Medicine, Technical University of Munich, Munich, Germany\and 
Institute for AI in Medicine (IKIM), University Medicine Essen, Essen, Germany\and
German Cancer Consortium (DKTK), partner site Essen\and
Pattern Analysis and Learning Group, Department of Radiation Oncology, Heidelberg University Hospital, Heidelberg, Germany\and
National Center for Tumor Diseases (NCT), NCT Heidelberg, a partnership between DKFZ and university medical center Heidelberg
\email{maximilian.fischer@dkfz-heidelberg.de}
}
\maketitle              
\begin{abstract}
Processing histopathological Whole Slide Images (WSI) leads to massive storage requirements for clinics worldwide. 
Even after lossy image compression during image acquisition, additional lossy compression is frequently possible without substantially affecting the performance of deep learning-based (DL) downstream tasks. 
In this paper, we show that the commonly used JPEG algorithm is not best suited for further compression and we propose Stain Quantized Latent Compression (\textit{SQLC}), a novel DL based histopathology data compression approach. \textit{SQLC} compresses staining and RGB channels before passing it through a compression autoencoder (\textit{CAE}) in order to obtain quantized latent representations for maximizing the compression. We show that our approach yields superior performance in a classification downstream task, compared to traditional approaches like JPEG, while image quality metrics like the Multi-Scale Structural Similarity Index (MS-SSIM) is largely preserved. Our method is \href{https://anonymous.4open.science/r/SQLC-2E8C/}{online} available.

\end{abstract}
\section{Introduction}
\label{sec:intro}
Storing digital histopathological Whole Slide Images (WSI) leads to massive storage requirements for clinics all over the world. Retaining digital WSI requires large amounts of storage, which is expensive and thus a major burden for the large scale uptake of digital pathology. A common setting across WSI vendors is to use the lossy JPEG compression scheme with a quality factor of 80 during image acquisition for file size reduction \cite{80QualityFactorAcquisition}. For DL applications, research has shown that even further lossy compression beyond the standard one during image acquisition is often possible without heavily affecting the performance of DL algorithms \cite{chen_quantitative_2020,doyle_evaluation_2010,ghazvinian_zanjani_impact_2019}, which is beneficial to further reduce the still very large filesize of WSI images. \\
Prior research has mainly focused on the impacts of the JPEG algorithm for further compression. However, it is important to note, that JPEG was initially created 50 years ago for natural images with the goal to compress images without affecting the perceptual image quality for humans. As a result, neither the natural image domain nor the solely human-centric aspect of JPEG seems ideally suited for various WSI applications, including DL-based WSI analysis. \\
Considering the importance of the issue that data compression addresses in the field of engineering, it is well-described and thoroughly researched. Several compression strategies, which consistently surpass JPEG in terms of compression ratio and image quality have been developed and some of the most promising methods are based on compression autoencoders \cite{HierarchicalPriorMethod,NoiseaddingAndFactorizedPriorModel,TheisPioneering}. Since these approaches are relatively new, only a few compression schemes consider the particularities of the WSI domain for DL-based compression \cite{TellezNeuralImageCompressionPatho,CiompiNeuralImageCompression} and these methods mainly address the dimensionality problem to process WSI slides as a whole through dimensionality reduction to fit them to conventional GPUs, which is not equivalent to data compression \cite{HierarchicalPriorMethod}. \\
In this paper, we propose \textit{Stain Quantized Latent Compression (SQLC)}, a novel deep learning framework, specialized for the lossy compression of WSI data. Similar to conventional schemes, our approach also first transfers the image domain into another better-suited domain, where quantization is applied. But in contrast to conventional linear schemes, our method is based on a non-linear transformation that is specifically tailored to pathological images. Our method builds upon an existing image compression method \cite{NoiseaddingAndFactorizedPriorModel} and quantizes the latent representation of the combined RGB and staining channels of WSI. Our contributions can be summarized as follows: (i) We show adverse effects for the further lossy JPEG compression of WSI data for ImageNet pre-trained models during a downstream task. (ii) We propose a novel approach to leverage the potential of existing DL-based compression schemes for WSI data. (iii) We provide exhaustive testing of the proposed method and compare several compression schemes for WSI data. Our results show that the proposed \textit{SQLC} method outperforms JPEG for compression ratios and systematically yields higher classification accuracies in downstream tasks. 

\section{Methods}
\subsection{SQLC Framework}

\begin{figure}[!htb]
    \centering
    \includegraphics[width=\linewidth,trim={0cm 3cm 0cm 2cm},clip]{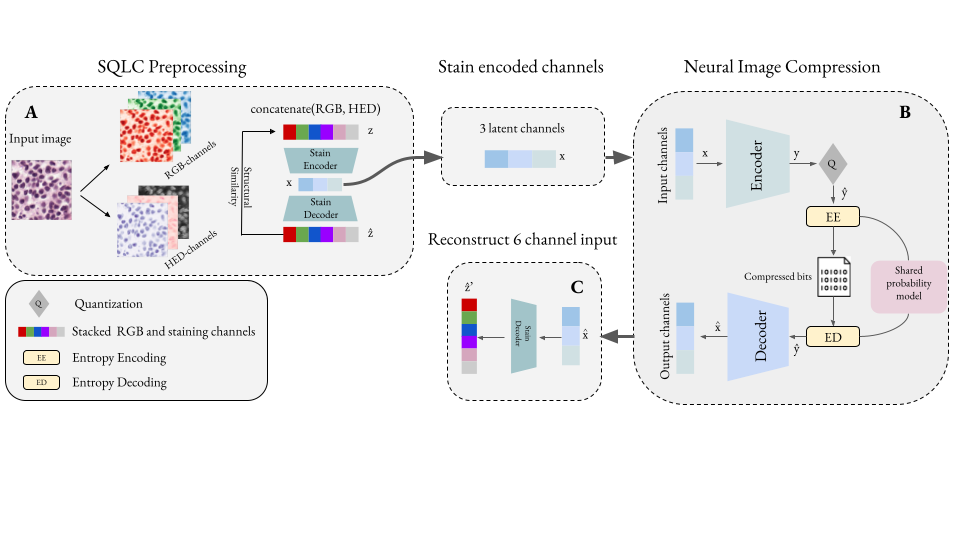}
    \caption{Proposed \textit{SQLC} method. A: Preprocessing for the fusion of RGB image features and staining features HED. B: Lossy compression of the latent representations, derived from part A. C: Finetuning of the decoder part of A to reconstruct the color channels from the quantized latents.}     
    \label{fig:MethodOverview}
\end{figure}

A summary of the \textit{SQLC} Framework can be found in figure \ref{fig:MethodOverview}. As an initial step, staining deconvolution is applied and the resulting staining channels are concatenated together with the three RGB-channels of an input image $z$. An auto-encoder generates a compressed version of the RGB- and staining-channels $x$. We refer to this autoencoder as stain encoder (SE). In part \textit{B} of the figure, the latent representation of the SE model are further processed for the lossy compression by a compression autoencoder, which yields $\hat{x}$. In the last step, depicted by part \textit{C} in the figure, the decoder part of the SE model reconstructs the 6-channel input (RGB+HED) $\hat{z}'$. Further details on the method can be found in \ref{StainDeconvSection} 

\subsection{Neural Image Compression (\textit{NIC})}\label{NICM}
Implicitly, any sort of data loss is inevitable for lossy compression, which is for discrete images achieved by quantization. The images themselves are not quantized, though; rather, they are first translated to a domain that is more suitable for quantization. Linear projections, including the JPEG algorithm's discrete cosine transformation, are constrained in their search for appropriate quantization domains, whereas neural networks are able to learn more complex non-linear transformations~\cite{TheisPioneering,NoiseaddingAndFactorizedPriorModel,HierarchicalPriorMethod}.\\
The most common \textit{NIC} schemes are based on variational autoencoders (VAE) as introduced in \cite{TheisPioneering} and differ from conventional VAEs by quantizing the latents and generally consist of three parts: (i) An encoder network $f_{\psi}$ learning non-linear functions, that generate a lower dimensional latent representation $y$ of the input data $x$. (ii) A distribution of the probabilities of occurrence $p_{\psi}$ of the quantized latents, necessary for an entropy-based coding algorithm for the lossless encoding of the quantized latents. (iii) A decoder $g_{\psi}$ that learns to decode the latents and to generate the reconstructed image $\hat{x}$.\\
This formalism demands two training requirements: First, the distortion $d$ between the original image $x$ and the reconstructed image $\hat{x}$, and second the length of the expected bitstream $r$ of the compressed output shall be minimal. Combining the aforementioned aspects results in the following bitlength-distortion optimization \cite{HierarchicalPriorMethod}: 
{\small
\begin{equation}
\label{eqn:LossFunction}
\splitfrac{
\mathcal{L} = r+\lambda\cdot d = \underbrace{\mathbb{E}_{x \sim p(x)} \lbrack -log_{2} p_{\psi}(\lbrack f_{\psi} (x) \rbrack) \rbrack}_\text{Bitrate}} {+ \underbrace{\lambda \cdot \mathbb{E}_{x \sim p(x)} \lbrack d(x,g_{\psi}(\lbrack f_{\psi} (x)\rbrack))\rbrack}_\text{image similarity}}
\end{equation}}
with the encoder $f_{\psi}$, the decoder $g_{\psi}$ and the probability model $p_{\psi}$. 
The image similarity can be comparably easily determined and optimized by common measures like the Multi Scale Structural Similarity Measure (MS-SSIM) \cite{MSSSIM}. On the other hand, quantization is a central element to minimize the length of the bitstream ($r$), which is unfortunately not differentiable per se. Hard quantization into non-continuous values, e.g. rounding, is not differentiable and thus a common mitigation strategy is to add a uniform noise distribution to relax the quantization problem and to approximate quantization during test time \cite{NoiseaddingAndFactorizedPriorModel}. 

\subsection{Stain Quantized Latent Compression (SQLC)}\label{StainDeconvSection}
In contrast to previous work, we present a method for lossy compression, specifically designed for WSI data. Our approach incorporates features from the RGB- and staining-domain of WSI files and processes them jointly. This is motivated by studies like \cite{combinedcolorchannelsGood,StainDeconvGoodDownstream}, where it has been shown that combined color channels are beneficial for the performance of DL models. Unlike standard color transfers such as those based on the HSV color space, stain deconvolution involves a more intricate analysis, considering the specific biochemical interactions of the staining process.\\
Figure \ref{fig:MethodOverview} shows a schematic overview of the proposed method. First, the staining components of the HE-stained WSI are separated into the Hematoxylin, Eosin and DAB components, following the approach of \cite{Vahadane}. The three staining channels are concatenated with the three original RGB channels, resulting in $z$. Second, information from both domains is fused, using an autoencoder model called stain encoder (SE) model. During training of the SE model, the image similarity between $z$ and the reconstruction $\hat{z}$ shall be maximized, which is determined by the MS-SSIM metric. The latent representation $x$ is then used as input for a subsequent lossy compression performed by a Neural Image Compression Model (NICM) based on \cite{NoiseaddingAndFactorizedPriorModel}. 
The NICM model reconstructs the input image as $\hat{x}$, and the tradeoff between higher image distortions between $x$ and $\hat{x}$ for low bit rates and vice versa is described with \ref{eqn:LossFunction}.\\
However, the reconstruction of $\hat{z}$ from $\hat{x}$ using the SE decoder is no longer possible, since $\hat{x}$ is the quantized representation of $x$. To address this issue, an additional Euclidean distance between $\hat{z}$ and $\hat{z}'$ is added to the loss function from \ref{eqn:LossFunction}, when training the NICM model. Here, $\hat{z}$ is generated by the SE decoder from $x$, while $\hat{z}'$ is generated from the reconstruction of $\hat{x}$ by the NICM method. This ensures a stable reconstruction process. The training process of the SQLC method involves three main steps: (i) Training the SE model to fuse features from both RGB and HED domains. (ii) Training the NICM on the latent representations from the SE encoder to perform effectively the lossy compression. (iii) Preserving the decoding capabilities of the SE decoder for quantized latents ($\hat{x}$) by adding the distance loss between $\hat{z}$ and $\hat{z}'$. Throughout the training process of the NICM, the weights of the SE model are kept frozen.

\section{Experiments}
\subsection{Dataset}
For the experiments, two different datasets are collected: One dataset that is used for training the compression model and one dataset that is used to evaluate the performance for a downstream task. 
We train the data compression task in an unsupervised scheme and use HED-stained breast and colon WSI at $40\times$ magnification from \cite{BreaKHis,LungColonData,KatherDataSet}. For the downstream task in the histopathology domain, we use the Camelyon16~\cite{ehteshami_bejnordi_diagnostic_2017} WSI dataset. The dataset contains lymph-node WSI and the 270 training and 130 test subjects are divided on a subject level into the classes metastasis and non-metastasis.

\subsection{Neural Image compression models}
\begin{figure}[h!]
    \centering
    \includegraphics[width=\linewidth,trim={3.5cm 7cm 10cm 6cm},clip]{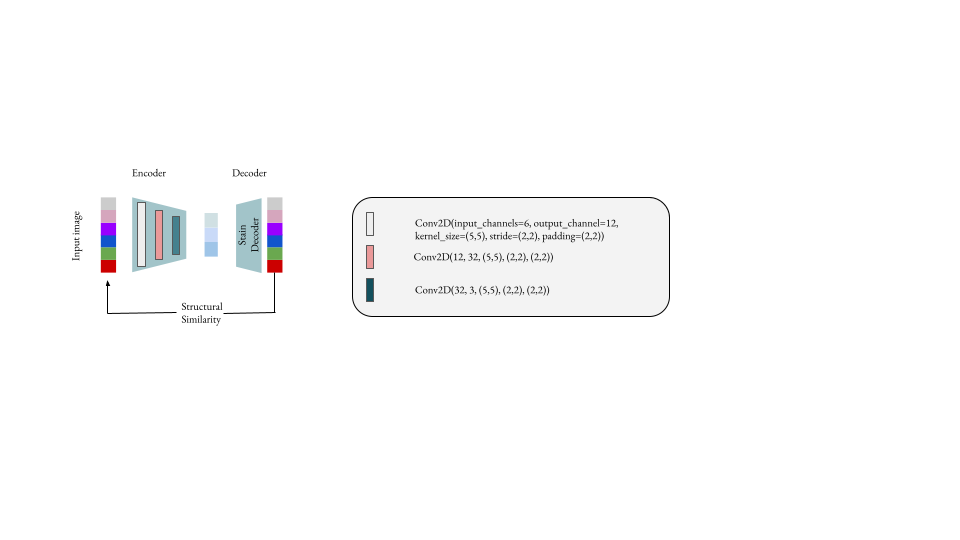}
    \caption{Architecture of the SE model.}
    \label{Figure:SEmodel}
\end{figure}

As a baseline neural image compression model, we implement the "factorized prior" model described in \cite{NoiseaddingAndFactorizedPriorModel} and we use the framework of \cite{begaint2020compressai} to implement the model in Pytorch \cite{Pytorch}. In the baseline setting, we train the model for 1000 epochs with a learning rate of $0.0001$, which is reduced on a plateau, where we sample values for $\lambda$ within the range of $[0,...,1.6]$, which are referred to as quality factors. As augmentations, we use random center crops with a patch size of $224 \times 224$ pixels and random horizontal and vertical flips. We refer to this method as \textit{baseline NICM}. 
As a variation of this approach, we also include stain augmentations during training, referred to as \textit{augmented NICM}. Here we perform a stain deconvolution during training with \cite{Vahadane} and augment both staining channels with the parameters $\alpha=[-0.05,0.05]$ and $\beta=[-0.2,0.2]$.\\
For the \textit{SQLC} framework, we extend the \textit{baseline NICM} by the additional convolutional autoencoder (SE model). Details on the architecture can be found in figure \ref{Figure:SEmodel}. 

For training of the SE model, we use patches of $224 \times 224$ pixels and train without further augmentations besides random horizontal and vertical flips. As a loss function, we use the channel-wise MS-SSIM and we train the model with a cosine annealing learning rate with the initial value of $0.01$ for 40 epochs and a batch size of $32$. 

\subsection{Experimental setup}
To quantify the effects of \textit{NIC} schemes, we compare several compression schemes during a downstream task. As a downstream task, we train an ImageNet pre-trained ResNet18 \cite{ResNets} model on the data of \cite{ehteshami_bejnordi_diagnostic_2017} to distinguish between cancerous and non-cancerous patches during four-fold cross-validation, where we use the Monai \cite{monai} framework with a binary cross-entropy loss function. During inference, we apply several compression schemes on the test data and evaluate the performance of the pre-trained model. As a lower baseline, we consider the further compression of the inference data with the lossy JPEG algorithm, which we generate with the python package PIL\footnote{\url{https://pypi.org/project/Pillow/}} for multiple quality factors from $[1,...,95]$. To determine the impact of the proposed \textit{SQLC} method, we compare it against the compression with \textit{baseline NICM, augmented NICM}, and \textit{SQLC} also for varying degrees of quantization. As a further ablation study on the impact of the additional overhead of the \textit{SQLC} method, we also evaluate a scenario where the encoder part of the \textit{baseline NICM} is modified to compress the 6 RGB- and HED-channels directly. This setting is described with \textit{NICM$_{6}$}.

\section{Results} \label{sec:results}
\subsection{Training of compression models} \label{CompressionResults}
\begin{figure}[htb!]
    \centering
    \includegraphics[width=\linewidth]{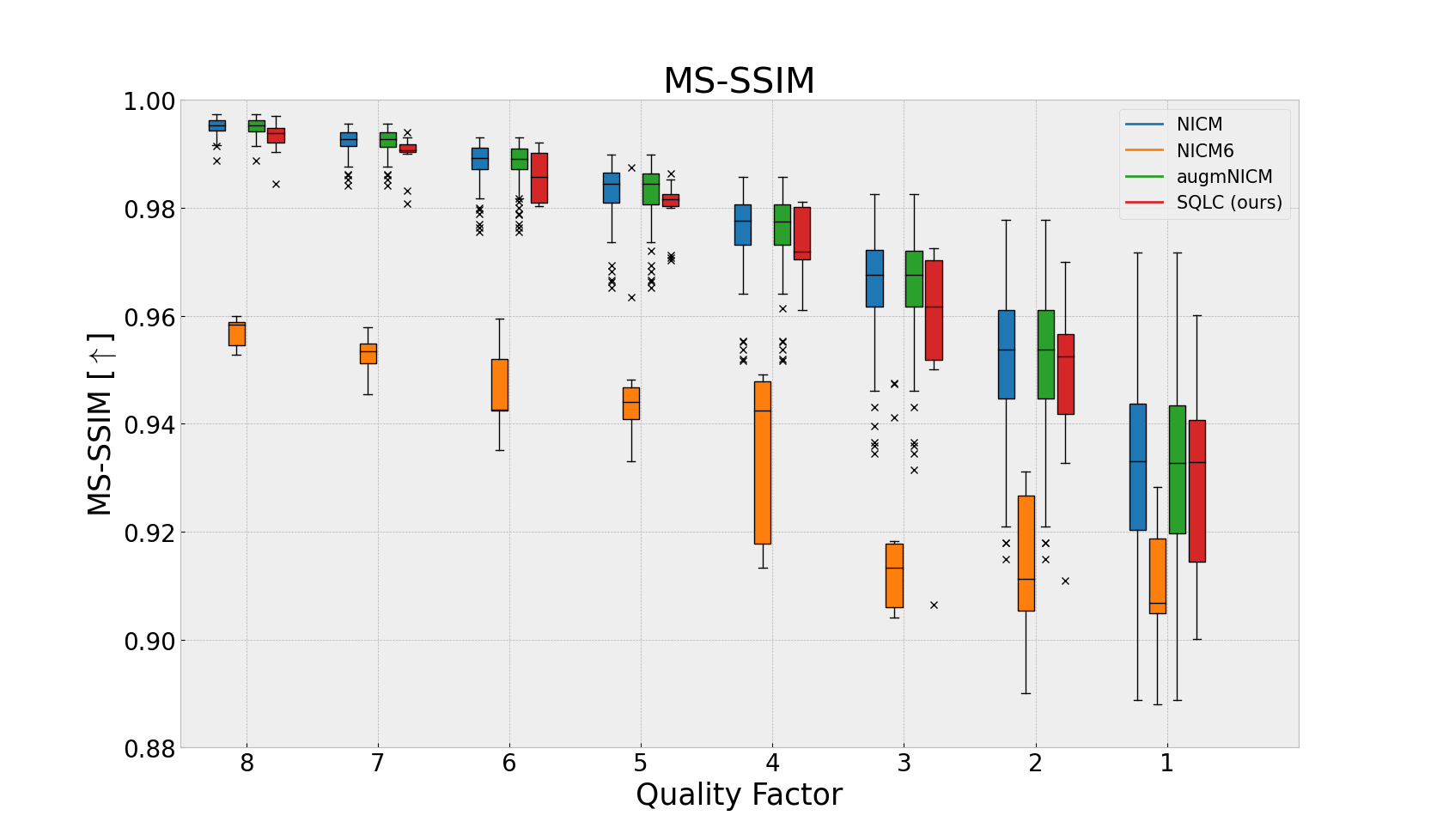}
    \caption{Perceptual Image quality score, evaluated with the MS-SSIM metric. The plot shows the achieved score for each compression scheme for the quality factors 8 - 1, which refer to increasing compression ratio.}
    \label{Figure:MSSSIMBoxplots}
\end{figure}

During training of lossy image compression models, the bit rate and the image distortion shall be kept minimal. Figure \ref{Figure:MSSSIMBoxplots} shows the rate-distortion performance of the trained \textit{NICM} models measured by the MS-SSIM metric. The first quality setting in the plot shows the baseline setting, which is besides minor deviations, consistent with the literature \cite{NoiseaddingAndFactorizedPriorModel,TheisPioneering,HierarchicalPriorMethod}. The plot shows that most dl-based compression schemes, including our SQLC scheme maintain reasonable amounts of image quality, besides the NICM$_{6}$ model. Reconstructions for the evaluation of the perceptual image quality can be found in the supplementary material and \href{https://anonymous.4open.science/r/SQLC-2E8C/}{online}.

\begin{figure}[htb!]
    \centering
    \includegraphics[width=\linewidth,trim={1cm 10cm 5cm 1cm},clip]{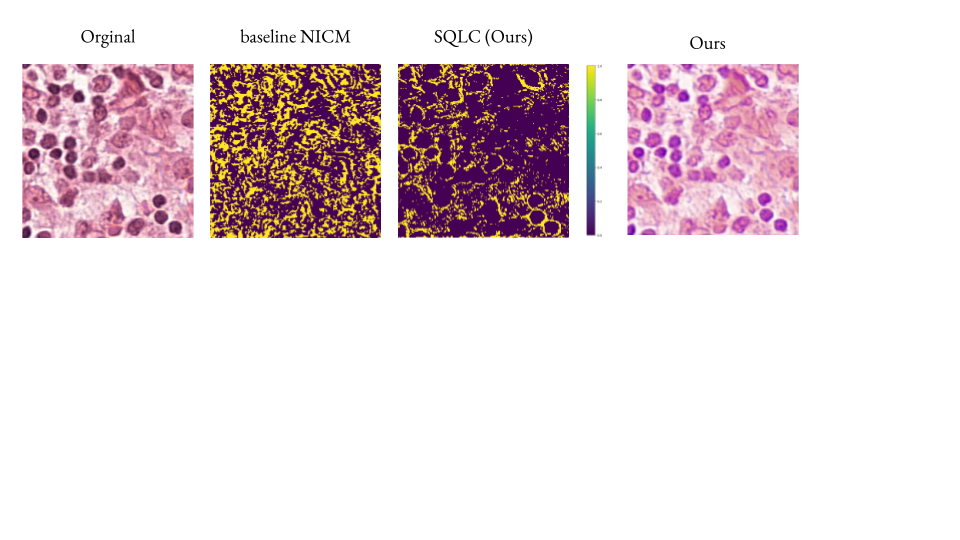}
    \caption{H-channel for the compression schemes. On the right side an example reconstruction from the \textit{SQLC} method is shown. The figure shows that the proposed method merely alters the information in the staining channels in contrast to other not finetuned compression schemes. }
    \label{Figure:HchannelDiff}
\end{figure}

Figure \ref{Figure:HchannelDiff} qualitatively shows the impact of the \textit{SQLC} method. The figure shows the difference in the Hematoxylin-channel for a sample image for a compression with the quality factor 4. For that visualization, first the original H-channel of a sample is extracted, subsequently the MSE error is calculated for the reference H-channel and the H-channels of the compression schemes baseline \textit{NICM} (2nd left) and  \textit{SQLC} (2nd right). The figure also shows a complete reconstruction after compressing it with our proposed method in the RGB-domain in the outer right image. Further difference images can be found in the supplementary material and \href{https://anonymous.4open.science/r/SQLC-2E8C/}{online}.

\subsection{Classification on compressed images} \label{InferenceResults}
For the downstream task, we classify tumor and non-tumor patches. Table \ref{Table:DownstreamResults} shows the AUC metrics for various lossy compression settings during inference. 
\begin{table}[]
\centering
\label{Table:DownstreamResults}
\caption{Mean and std value of the AUC metric for the classification task with various compression techniques. For visualization purposes, we only show quality factors $8,6,5,4,2,1$. The complete set of quality factors can be found in the supplementary material.}
\begin{tabular}{lcccccc}
\toprule
\multirow{2}{*}{Compression}  & \multicolumn{6}{c}{AUC at Compression of} \\ \cmidrule{2-7}
  & 1:3    & 1:12    & 1:22    & 1:34     & 1:56    & 1:73    \\ \midrule
JPEG            & 88.2±0.2 & 84.8±0.2 & 80.6±1.3 & 79.2±0.7  & 79.3±1.4    & --    \\
NICM            & \textbf{89.5}±0.1 & 87.3±0.1 & 87.2±0.1 & 86.6±0.3  & 85.2±0.3 & 84.7±0.3 \\
NICM$_{6}$      &76.0±0.1 & 75.2±0.1 & 76.0±0.1&74.0±0.1&64.9±0.1& 63.9±0.1\\
augm. NICM      & 87.3±0.2 & 87.2±0.1 & 86.9±0.2 & 86.2±0.1  & 85.0±0.1 & 84.5±0.1 \\
SQLC (ours)     & 89.5±0.2 & \textbf{89.5}±0.2 & \textbf{89.1}±0.1 & \textbf{88.5}±0.1  & \textbf{86.9}±0.3 & \textbf{86.5}±0.2\\
\bottomrule
\end{tabular}
\end{table}
The table shows that with JPEG compression, high compression ratios result in a sharp drop in classification accuracy. In contrast to this, the AUC value for the compression with \textit{baseline NICM} and the \textit{augmented NICM} only slightly decreases. The table shows that the proposed SQLC technique outperforms the other approaches, especially for higher compression ratios.

\section{Discussion}
This paper presents a novel dl-based lossy image compression scheme, which is specifically tailored for applications in digital pathology and very high compression ratios. Our approach enables further compression of pathology images while maintaining high classification accuracies. While deep neural networks are theoretically also capable of compressing the combined staining and RGB channels directly into one embedding, without the SE model's inductive bias, in our approach, we split encoding and decoding into two models (the SE and the NICM). For image compression in general and for digital pathology in particular, encoding and decoding of images is often done on different clients with different hardware resources. In clinical settings, encoding of images is usually done on the microscope or on powerful servers in the hospital, while images are decoded on lightweight browser-based image viewing infrastructure components. Consequently, it can be considerably easier to implement a modular strategy that can be built upon several backbone compression approaches. We also show that shallow encoder models, like the NICM method, are not capable of learning meaningful representations of the 6 input channels directly. With the qualitative figures in section \ref{sec:results}, we explicitly show how our method differs from other dl-based compression schemes and leaves the important staining channels unaffected by alterations in the image quality. While we demonstrate this with a classification task, we hypothesize that this also holds true for further tasks. In future experiments, we plan to investigate further tasks and also specialized stainings like immunohistochemical staining. Higher compression ratios for reducing the tremendous storage requirements for pathology laboratories worldwide are of increasing need. Compared to other compression schemes, our approach enables very high compression ratios without significantly affecting the downstream task performance. In future experiments, we also plan to investigate specialized image decoding algorithms, like the one presented in \cite{TheisDiffusion}.

\newpage

\bibliographystyle{splncs04}
\bibliography{BibFile}

\begin{thebibliography}{10}
\providecommand{\url}[1]{\texttt{#1}}
\providecommand{\urlprefix}{URL }
\providecommand{\doi}[1]{https://doi.org/#1}

\bibitem{CiompiNeuralImageCompression}
Aswolinskiy, W., Tellez, D., Raya, G., et~al.: Neural image compression for non-small cell lung cancer subtype classification in h\&e stained whole-slide images. In: Tomaszewski, J.E., Ward, A.D. (eds.) Medical Imaging 2021: Digital Pathology. vol. 11603, p. 1160304. International Society for Optics and Photonics, SPIE (2021). \doi{10.1117/12.2581943}

\bibitem{NoiseaddingAndFactorizedPriorModel}
Ball{\'e}, J., Laparra, V., Simoncelli, E.P.: End-to-end optimized image compression. In: International Conference on Learning Representations (2017), \url{https://openreview.net/forum?id=rJxdQ3jeg}

\bibitem{begaint2020compressai}
B{\'e}gaint, J., Racap{\'e}, F., Feltman, S., Pushparaja, A.: Compressai: a pytorch library and evaluation platform for end-to-end compression research. arXiv preprint arXiv:2011.03029  (2020)

\bibitem{LungColonData}
Borkowski, A.A., Bui, M.M., Thomas, L.B., Wilson, C.P., DeLand, L.A., Mastorides, S.M.: Lung and colon cancer histopathological image dataset (lc25000) (2019). \doi{10.48550/ARXIV.1912.12142}

\bibitem{monai}
Cardoso, M.J., Li, W., Brown, R., et~al.: Monai: An open-source framework for deep learning in healthcare (2022). \doi{10.48550/ARXIV.2211.02701}

\bibitem{chen_quantitative_2020}
Chen, Y., Janowczyk, A., Madabhushi, A.: Quantitative assessment of the effects of compression on deep learning in digital pathology image analysis. {JCO} clinical cancer informatics  \textbf{4},  221--233 (2020). \doi{10.1200/CCI.19.00068}

\bibitem{doyle_evaluation_2010}
Doyle, S., Monaco, J., Madabhushi, A., et~al.: Evaluation of effects of {JPEG}2000 compression on a computer-aided detection system for prostate cancer on digitized histopathology. In: 2010 {IEEE} International Symposium on Biomedical Imaging: From Nano to Macro. pp. 1313--1316. {IEEE} (2010). \doi{10.1109/ISBI.2010.5490238}

\bibitem{ehteshami_bejnordi_diagnostic_2017}
Ehteshami~Bejnordi, B., Veta, M., Johannes~van Diest, P., van Ginneken, B., Karssemeijer, N., Litjens, G., van~der Laak, J.A.W.M., the CAMELYON16~Consortium: Diagnostic assessment of deep learning algorithms for detection of lymph node metastases in women with breast cancer. {JAMA}  \textbf{318}(22), ~2199 (2017). \doi{10.1001/jama.2017.14585}

\bibitem{ghazvinian_zanjani_impact_2019}
Ghazvinian~Zanjani, F., Zinger, S., Piepers, B., Mahmoudpour, S., Schelkens, P.: Impact of {JPEG} 2000 compression on deep convolutional neural networks for metastatic cancer detection in histopathological images. Journal of Medical Imaging  \textbf{6}(2), ~1 (2019). \doi{10.1117/1.JMI.6.2.027501}

\bibitem{combinedcolorchannelsGood}
Gowda, S.N., Yuan, C.: Colornet: Investigating the importance of color spaces for image classification. In: Jawahar, C., Li, H., Mori, G., Schindler, K. (eds.) Computer Vision -- ACCV 2018. pp. 581--596. Springer International Publishing, Cham (2019)

\bibitem{ResNets}
He, K., Zhang, X., Ren, S., Sun, J.: Deep residual learning for image recognition. {arXiv}:1512.03385 [cs]  (2015), \url{http://arxiv.org/abs/1512.03385}

\bibitem{StainDeconvGoodDownstream}
Janowczyk, A., Madabhushi, A.: Deep learning for digital pathology image analysis: A comprehensive tutorial with selected use cases. Journal of Pathology Informatics  \textbf{7}(1), ~29 (2016). \doi{https://doi.org/10.4103/2153-3539.186902}

\bibitem{KatherDataSet}
{Kather}, J.N., {Weis}, C.A., {Bianconi}, F., {Melchers}, S.M., {Schad}, L.R., {Gaiser}, T., {Marx}, A., {Z{\"o}llner}, F.G.: {Multi-class texture analysis in colorectal cancer histology}. Scientific Reports  \textbf{6},  27988 (Jun 2016). \doi{10.1038/srep27988}

\bibitem{HierarchicalPriorMethod}
Minnen, D., Ball\'{e}, J., Toderici, G.D.: Joint autoregressive and hierarchical priors for learned image compression. In: Advances in Neural Information Processing Systems. vol.~31. Curran Associates, Inc. (2018), \url{https://proceedings.neurips.cc/paper/2018/file/53edebc543333dfbf7c5933af792c9c4-Paper.pdf}

\bibitem{Pytorch}
Paszke, A., Gross, S., Massa, F., et~al.: Pytorch: An imperative style, high-performance deep learning library. In: Wallach, H., Larochelle, H., Beygelzimer, A., d\textquotesingle Alch\'{e}-Buc, F., Fox, E., Garnett, R. (eds.) Advances in Neural Information Processing Systems. vol.~32. Curran Associates, Inc. (2019), \url{https://proceedings.neurips.cc/paper/2019/file/bdbca288fee7f92f2bfa9f7012727740-Paper.pdf}

\bibitem{BreaKHis}
Spanhol, F.A., Oliveira, L.S., Petitjean, C., Heutte, L.: A dataset for breast cancer histopathological image classification. IEEE Transactions on Biomedical Engineering  \textbf{63}(7),  1455--1462 (2016). \doi{10.1109/TBME.2015.2496264}

\bibitem{80QualityFactorAcquisition}
Stathonikos, N., Veta, M., Huisman, A., {van Diest}, P.J.: Going fully digital: Perspective of a dutch academic pathology lab. Journal of Pathology Informatics  \textbf{4}(1), ~15 (2013). \doi{https://doi.org/10.4103/2153-3539.114206}

\bibitem{TellezNeuralImageCompressionPatho}
Tellez, D., Litjens, G., van~der Laak, J., Ciompi, F.: Neural image compression for gigapixel histopathology image analysis. IEEE Transactions on Pattern Analysis and Machine Intelligence  \textbf{43}(2),  567--578 (2021). \doi{10.1109/TPAMI.2019.2936841}

\bibitem{TheisDiffusion}
Theis, L., Salimans, T., Hoffman, M.D., Mentzer, F.: Lossy compression with gaussian diffusion (2022), \url{https://arxiv.org/abs/2206.08889}, arXiv:2206.08889

\bibitem{TheisPioneering}
Theis, L., Shi, W., Cunningham, A., Husz{\'a}r, F.: Lossy image compression with compressive autoencoders. In: International Conference on Learning Representations (2017), \url{https://openreview.net/forum?id=rJiNwv9gg}

\bibitem{Vahadane}
Vahadane, A., Peng, T., Sethi, A., et~al.: Structure-preserving color normalization and sparse stain separation for histological images. IEEE Transactions on Medical Imaging  \textbf{35}(8),  1962--1971 (2016). \doi{10.1109/TMI.2016.2529665}

\bibitem{MSSSIM}
Wang, Z., Simoncelli, E., Bovik, A.: Multiscale structural similarity for image quality assessment. vol.~2, pp. 1398 -- 1402 Vol.2 (12 2003). \doi{10.1109/ACSSC.2003.1292216}

\end{thebibliography}
\newpage

\section{Appendix}

\begin{table}[]
\centering
\caption{Description of the dataset.}
\begin{tabular}{cccc}
Dataset      & BreaKHis & Colon1 & Colon2 \\ \hline
Source       & [16]         & [4]       &  [13]       \\
Images       &   1639         &500        &10        \\
SampleSize   &700x460          &768x768        &5000x5000        \\
Tiles        &10158          &10005        &4840        \\

\begin{tabular}{l}
Image
\end{tabular} &   \includegraphics[width=20mm,height=20mm]{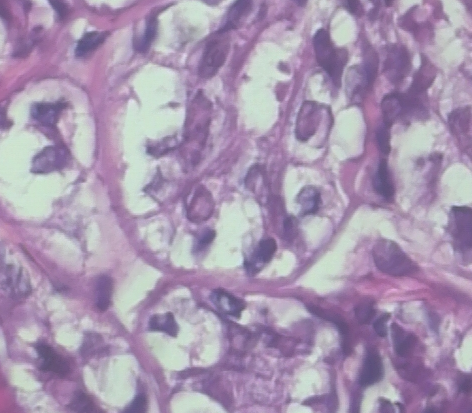}       &\includegraphics[width=20mm,height=20mm]{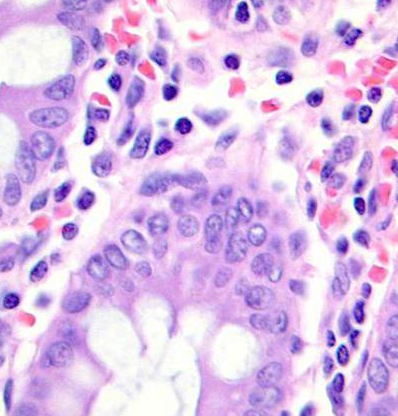}        &       \includegraphics[width=20mm,height=20mm]{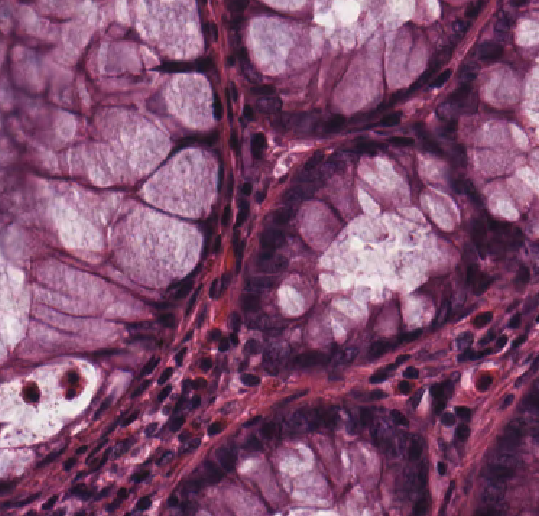}\\
&&&\\
\end{tabular}
\begin{tabular}{ccccc}
\hline
Camelyon16  & \multicolumn{2}{c}{Train} & \multicolumn{2}{c}{Test} \\
          & tumor      & no tumor     & tumor     & no tumor     \\ \hline
WSI Files & 110           & 160             &48           &  82            \\
Tiles     & 28735           &79418              &11097           &40500  \\

\end{tabular}
\end{table}

\begin{figure}[h!]
    \centering
    \includegraphics[width=\textwidth,trim=10pt 180pt 10pt 10pt, clip]{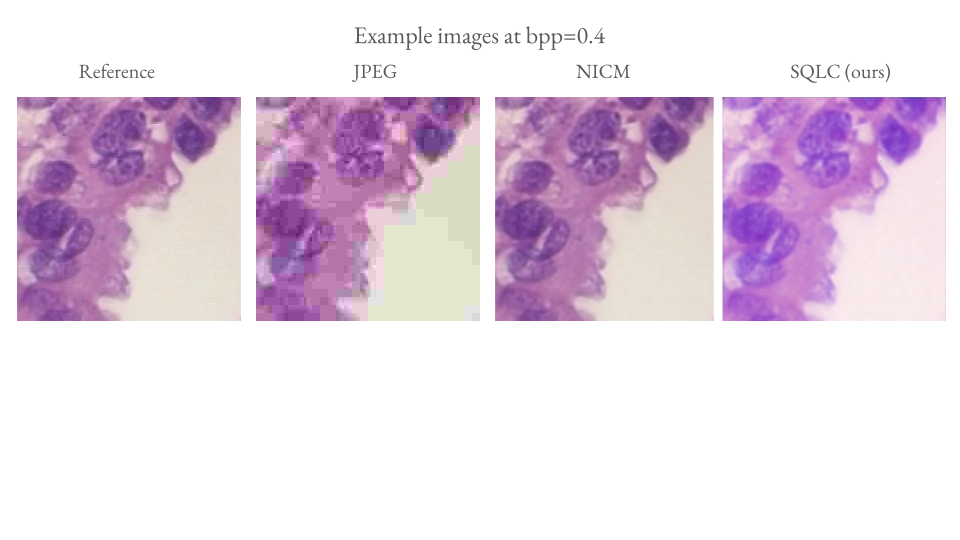}
    \caption{Visualizations of reconstructed images with some of the compared compression schemes.}
    \label{Figure:LatentsAndAUC}
\end{figure}

\begin{table}[]
\centering
\rotatebox{90}{%
\begin{tabular}{lcccccccc}
\toprule
\multirow{2}{*}{\textbf{Compression}}  & \multicolumn{6}{c}{\textbf{AUC at Compression of}} \\ \cmidrule{2-7}
  & \textbf{1:3}  & \textbf{1:9}  & \textbf{1:12}    & \textbf{1:22}    & \textbf{1:34}  & \textbf{1:44}   & \textbf{1:56}    & \textbf{1:73}    \\ \midrule
JPEG            & 88.2±0.2 &86.0±0.2 & 84.8±0.2 & 80.6±1.3 & 79.2±0.7 &79.9±0.2  & 79.3±1.4    & --    \\
NICM            & \textbf{89.5}±0.1 &88.5±0.2 & 87.3±0.1 & 87.2±0.1 & 86.6±0.3 &85.8±0.3  & 85.2±0.3 & 84.7±0.3 \\
NICM$_{6}$      &76.0±0.1 &75.7±0.2 & 75.2±0.1 & 76.0±0.1&74.0±0.1 &69.2±0.2 &64.9±0.1& 63.9±0.1\\
augm. NICM      & 87.3±0.2 &86.2±0.9 & 87.2±0.1 & 86.9±0.2 & 86.2±0.1  &85.8±0.2 & 85.0±0.1 & 84.5±0.1 \\
SQLC (ours)     & 89.5±0.2 &\textbf{89.4±0.2} & \textbf{89.5}±0.2 & \textbf{89.1}±0.1 & \textbf{88.5}±0.1 &\textbf{87.4±0.2}  & \textbf{86.9}±0.3 & \textbf{86.5}±0.2\\
\bottomrule
\end{tabular}%
}
\caption{Mean and std value of the AUC metric.}
\label{Table:DownstreamResults}
\end{table}

\end{document}